\documentclass[aps,pre,amsmath,amssymb,twocolumn,groupedaddress,showpacs]{revtex4}
\usepackage{graphicx}
\usepackage[utf8]{inputenc} 

\usepackage{epsf}

\newcommand{\be}{\begin{equation}}
\newcommand{\ee}{\end{equation}}
\newcommand{\ba}{\begin{eqnarray}}
\newcommand{\ea}{\end{eqnarray}}
\newcommand{\baa}{\begin{eqnarray}}
\newcommand{\eaa}{\end{eqnarray}}
\newcommand{\ed}{\end{document}}

\newcommand{\re}[1]{(\ref{#1})}

\renewcommand{\baselinestretch}{1.2}
\date{\today}
\begin{document}%\large
\title{Quantum Transport in Ladder-Type Networks: Role of nonlinearity, topology and spin}
\author{K. Nakamura$^{1,2}$}
\email{nakamura@a-phys.eng.osaka-cu.ac.jp}
\author{D. Matrasulov$^1$}
\author{G. Milibaeva$^1$}
\author{J. Yusupov$^1$}
\author{U. Salomov$^1$}
\author{T.Ohta$^2$}
\author{M.Miyamoto$^2$}
\affiliation{
$^1$Heat Physics Department of the Uzbek Academy of Sciences,
28 Katartal Street, 100135 Tashkent, Uzbekistan
\\
$^2$Department of Applied Physics,
Osaka City University,
Osaka 558-8585, Japan
}

%\author{K.Nakamura, D. Matrasulov, G. Milibaeva, J. Yusupov and U. Salomov
%\\Heat Physics Department of the Uzbek Academy of Sciences\\
%28 Katartal St.,100135 Tashkent, Uzbekistan\\
%T.Ohta and M.Miyamoto\\
%Department of Applied Physics,\\Osaka City University, Osaka 558-8585, Japan\\
%}

\begin{abstract}
We investigate quantum transport of electrons, phase solitons, etc. through mesoscopic networks of zero-dimensional quantum dots. Straight and circular ladders are chosen as networks with each coupled with three semi-infinite leads (with one incoming and the other two outgoing).
Two transmission probabilities (TPs) as a function of the incident energy $\varepsilon$ show a transition from anti-phase aperiodic to degenerate periodic spectra at the critical energy $\varepsilon_c$  which is determined by a bifurcation point of the bulk energy dispersions. TPs of the circular ladder depend only on the parity of the winding number.   
Introduction of a single missing bond (MB) or missing step doubles the period of the periodic spectra at $\varepsilon>\varepsilon_c$ . Shift of the MB by lattice constant results in a striking switching effect at $\varepsilon<\varepsilon_c$.
In the presence of the electric-field induced spin-orbit interaction (SOI), an obvious spin filtering  occurs against the spin-unpolarized injection. Against the spin-polarized injection, on the other hand, the spin transport shows spin-flip (magnetization reversal) oscillations with respect to SOI.   We also show a role of soliton in the context of its transport through the ladder networks.
\end{abstract}
\maketitle

\section{Introduction}
Recently there has been a growing interest in  quantum transport in discrete physical systems characterized by networks with nontrivial topologies \cite{har,kott}. Those networks mimic networks of nonlinear waveguides and and optical fibers \cite{kiv} , Bose-Einstein condensates in optical lattices \cite{tro}, superconducting ladders of Josephson junctions \cite{bin}, double helix of DNA, etc. In these networks, their topology and the presence of a few embedded defects are expected to play a vital role in controlling the macroscopic quantum transport such as a switching of the network current. Here, a main interest lies in the networks connecting everywhere-discrete lattice points 
\cite{fla,abl} in contrast to another  topical works on quantum graphs which are composed of connected continuous linear segments of finite length \cite{kott}.

On the other hand, with  introduction of the nonlinearity to the time-dependent Schr\"odinger equation, the network provides a nice
playground where solitons propagate in a complicated way until escaping through the attached semi-infinite leads. There already exists
an accumulation of studies of the soliton propagation through the discrete chain, and its collision with small defect clusters \cite{bur}.
However, little work has been done on the soliton transport through the big networks with and without defects.

In this paper we investigate quantum transport of electrons or phase solitons  through mesoscopic networks of zero-dimensional
quantum dots. Typically, straight and circular ladders are chosen as model networks with each being coupled with three semi-infinite leads (with one
incoming and the other two outgoing).
In Section II , based on the discrete cubic nonlinear Schr\"odinger equation, we examine a fate of the soliton coming from the incoming lead and propagating through the above networks in a complicated way until escaping through the  three  semi-infinite leads. The two transmission probabilities (TPs) based on a soliton picture are evaluated and compared with the result of Landauer formula based on the (stationary and discrete) linear Schr\"odinger equation. The following Sections are based on the standard (linear) quantum mechanics. In Section III, TPs are explored as a function of the incident energy, and the characteristic features of the transmission spectra are found. In Section IV we shall elucidate a radical change of the transmission spectra by introducing a single defect bond into the network. The role of topology in the transport through the circular ladder is also studied in this Section.  Finally in Section V the electric-field induced spin-orbit interaction (i.e., Rashba interaction) is introduced to the network. Then we investigate the result of spin transport through the networks and indicate its role in magnetization oscillations and spin filtering. Summary and discussion are devoted to Section VI.

\section{Model Networks and Discrete Nonlinear Schr\"odinger equation}
As a challenge to analyze general big networks, we choose two type of networks, straight and circular ladders (see  Figs. \ref{fig1} and \ref{fig2}), which mimic Josephson junction or double helix of DNA.
Each system consists of an array of zero-dimensional quantum dots (i.e., lattice sites),  where central part represents a network and external three lines stand for the attached semi-infinite leads. All lattice points are numbered in the way given in Figs. \ref{fig1} and {\ref{fig2}. 
In Fig. \ref{fig1}, for example, the incoming lead (left) is connected with the ladder at the site $m$ and a pair of outgoing
leads (right) are connected with it at the sites $m+2n$ and $m+2n+1$. Suppressing three external leads, the ladder includes $2n+2$ lattice sites and $n-1$ steps (perpendicular to the ladder).
The wave function comes through the incoming
lead ($\Phi_{in}$), collides with the network, and is partly reflected through the incoming lead ($\Phi_{ref}$) and partly transmitted through two outgoing leads ($\Phi_{out1}, \Phi_{out2}$).
\begin{figure}[htb]
\centerline{\includegraphics[width=\columnwidth]{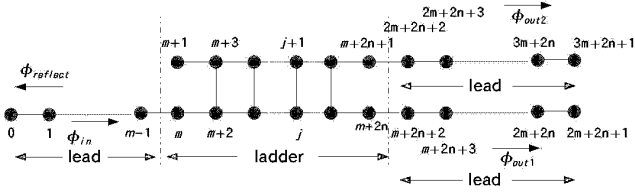}}
\caption{Straight ladder with 3 leads.} 
\label{fig1}
\end{figure}
\begin{figure}[htb]
\centerline{\includegraphics[width=\columnwidth]{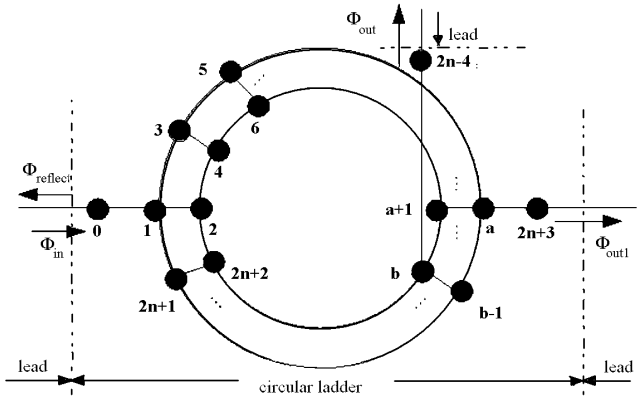}}
\caption{Circular ladder with 3 leads.} 
\label{fig2}
\end{figure}
Dynamics of a wave function in these open networks is described by discrete nonlinear Schr\"odinger equation (DNLSE),
\begin{equation}
 i\frac{\partial \Phi_j}{\partial t}=-\frac{1}{2}\sum_lA_{j,l}\Phi_l+\Lambda |\Phi_j|^2\Phi_j
\end{equation}
where $\Lambda$ represents the strength of cubic nonlinearity. $A_{ij}$ is adjacency matrix giving the topology of the network and is defined, in a suitable energy unit (say, $K$) by 
\begin{eqnarray} 
A_{j,l}= \begin{cases} 1 \quad & \text{ if\; $j$\; and\; $l$ \; are\; linked} \\
0  &  \text{otherwise} \\ \end{cases}
\end{eqnarray}
In the case of quantum dots with a common discrete level (CDL) for each, $\Phi_j(t)$ is the wave function of the $j$-th dot. The distances between linked lattice sites are fixed to a common value, say, $d$ with $d$ being of order of $10 \sim 100$ nm. $K$ stands for the tunneling matrix element between connected adjacent dots. CDL is chosen around Fermi energy and prescribed to zero energy. Time $t$ is in units of $\hbar/2K$ and $\Lambda=U/2K$ with $U$ the very weak Hartree term due to the electron-electron interaction.
Firstly we investigate the injection of a wave packet (WP) through the incoming lead, where DNLSE governs:
\begin{equation}
i\frac{\partial \Phi_j}{\partial t}=-\frac{1}{2}(\Phi_{j-1}+\Phi_{j+1})+\Lambda |\Phi_j|^2\Phi_j
\end{equation}
Consider, at $t=0$,  Gaussian WP centered at $\xi_0$, with initial 
momentum $k_0$ and width $\gamma_0$. 
In its discrete version the time-dependent WP can be written as 
\begin{equation}
\Phi_j(t)=\sqrt{N} \exp \left(\frac{-(j-\xi)^2}{\gamma^2}+ik(j-\xi)+i\frac{\delta}{2}(j-\xi)^2 \right) 
\end{equation}
where $\xi(t)$ and $\gamma(t)$, which are scaled by $d$, are time-dependent center of mass and width of WP, respectively. $k(t)$ and $\delta(t)$, which are scaled by $d^{-1}$ and $d^{-2}$, respectively, are the corresponding canonical-conjugate variables.

In the limit $\gamma d \gg d$, WP dynamics can be obtained from effective Lagrangian
\begin{equation}
L=k\dot\xi -\gamma^2\frac{\delta}{8}-\frac{\Lambda}{2\sqrt{\pi\gamma^2}}+\cos(k)e^{-\eta}
\end{equation}
from which we have the equations of motion for $\xi, k, \gamma$ and $\delta$.
%\begin{eqnarray}
%\dot{k} & = & 0 \nonumber \\
%\dot{\xi} & = & \sin{k} \, \cdot \, e^{-\eta} \nonumber \\ 
%\dot{\delta} & = & \cos{k} \Big(4 / \gamma^4-\delta^2 \Big) 
%e^{-\eta} + 2 \Lambda / \sqrt{\pi} \gamma^3 \nonumber \\ 
%\dot{\gamma} & = & \gamma \delta \cos{k} \, \cdot \, e^{-\eta}.\nonumber
%\end{eqnarray} 
In order to have a stable WP (soliton) on incoming leads it should be $\dot \gamma=\dot \delta=0$, from which it  follows \cite{mal,tro,bur}
\begin{equation}
\label{lambda_sol}
\Lambda_{sol} \approx  2 \sqrt{\pi} \frac{\mid \cos{k} \mid} 
{\gamma_0}.
\end{equation}
with $\frac{\pi}{2}\le k(=k_0) \le \pi$ and $\delta=0$.
\begin{figure}[htb]
\centerline{\includegraphics[width=\columnwidth]{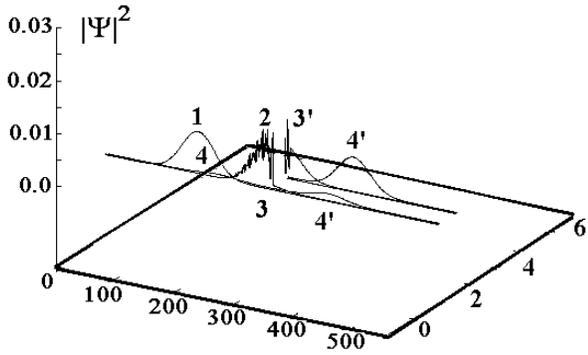}}
\caption{Soliton dynamics in straight ladder with 3 leads. Time evolution of the spatial distribution of the positive wave-function probability: $1\to 2\to (3,3')\to (4,4')$.
$k=\frac{5}{8}\pi$. Basal Lengths and wave number are scaled by $d$ and $d^{-1}$, respectively. Ladder steps are not depicted for simplicity.} 
\label{str_d}
\end{figure}
\begin{figure}[htb]
\centerline{\includegraphics[width=\columnwidth]{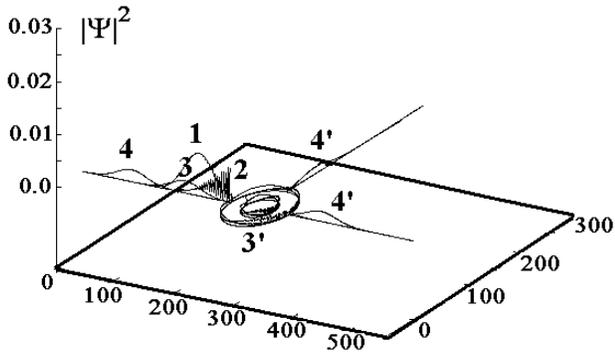}}
\caption{Soliton dynamics in circular ladder with 3 leads. Time evolution of the spatial distribution of the positive wave-function probability: $1\to 2\to (3,3')\to (4,4')$.
$k=\frac{3}{5}\pi$. The same notion on lengths, wave number and ladder steps holds as in Fig.\ref{str_d}.} 
\label{circ_d}
\end{figure}
Under this conditions we present the numerical results of soliton dynamics colliding with a network in Figs. \ref{str_d}} and {\ref{circ_d}.
Soliton propagates
through the incoming lead (marked as '1'), collides with network (marked as '2'), propagates through
network (marked as '3' and '3'') and is partially reflected through the incoming lead (marked as '4') and partially transmitted through two outgoing leads (marked as  '4'').
Transmission and reflection probabilities (TP and RP) at long enough time after collision with the network can be calculated as
\begin{eqnarray}
T_1&=&\sum_{j \in\;\text{outgoing lead 1}} |\Phi_j|^2 \nonumber \\
T_2&=&\sum_{j \in\;\text{outgoing lead 2}} |\Phi_j|^2 \nonumber \\
R&=&\sum_{j \in\;\text{incoming lead}} |\Phi_j|^2.
\label{nltr}
\end{eqnarray}
\begin{figure}[htb]
\centerline{\includegraphics[width=\columnwidth]{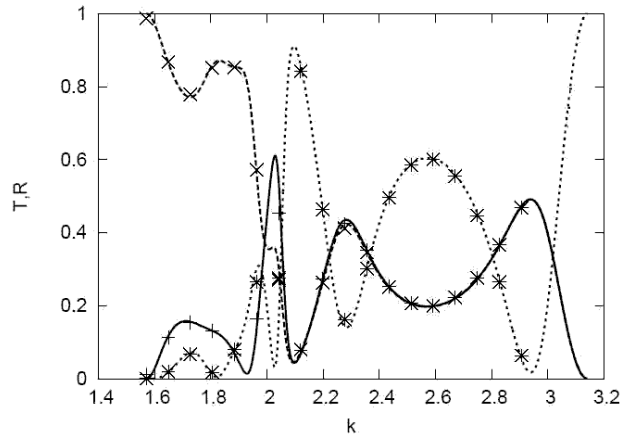}}
\caption{Comparison  $T_1, T_2$ and $R$ between Eq.~\re{nltr} with use of nonlinear dynamics of a soliton and Eq.~\re{landtr} in Landauer
formula for the time-independent linear Schr\"odinger equation. Number of steps in ladder is $n=10$. Solid line and '+' for $T_1$, dashed line and '$\times$' for $T_2$, and dotted line and '$\ast$' for $R$.} 
\label{comp}
\end{figure}
The result as a function of the incident wave number $k$ (scaled by $d^{-1}$) is shown in a set of symbols in Fig.~\ref{comp} in the case of the straight ladder with number of steps $n=10$ and length of each external lead $m=250$. Here initial width of wave packet $\gamma_0=50$ and initial center of mass $\xi_0=100$. We find the unitarity $T_1+T_2+R=1$ is always satisfied, namely no fraction of WP remains in the central 
network at long-enough time.

Also, we compare this result with the result based on Landauer formula \cite{but,dat} applied to  the time-independent linear Schr\"odinger equation 
for the ladder network with $N(=2n)$ lattice sites, which is connected with the semi-infinite incoming lead at '$0$' site and two semi-infinite outgoing leads at '$N+1$' and '$N+2$' sites.
In the latter approach, the outgoing wavefunction $\Psi=(\Phi_0,\Phi_1,\ldots,\Phi_{N+1},\Phi_{N+2})^T$ is determined by \cite{ando} 
\begin{equation}
\Psi=G\Psi_{in}
\label{land}
\end{equation}
against the incoming wave function $\Psi_{in}=(-K_s[F^{-1}(+)-F^{-1}(-)]\Phi_0(+),0,\ldots,0)^T$ with
$K_s$ and $F^{-1}(\pm)$ the tunneling and transfer matrices, respectively, in the leads.
$G$  is the Green function defined by
\begin{equation}
G=\frac{1}{E-\tilde H}.
\label{green}
\end{equation}
In  Eq.\re{green}, $\tilde H$ is the Hamiltonian which includes the interaction of the network with external leads \cite{ando,mir}:

\begin{equation}
\tilde H=\left(
\begin{array}{cccccc}
\tilde V_0 & K^{\ast}_{0,1}&0&\ldots&0&0\\
K_{0,1}& & & & \vdots&\vdots\\
0&&H&&K^{\ast}_{N-1,N+1}&0\\
\vdots&&&& 0&K^{\ast}_{N,N+2}\\
0&\ldots& K_{N-1,N+1}&0&\tilde V_{N+1}&0\\
0&\ldots& 0&K_{N,N+2}&0&\tilde V_{N+2}\end{array}\right)
\end{equation}
where $H$ is the unperturbed Hamiltonian. $\tilde V_0,\tilde V_{N+1},\tilde V_{N+2}$ and $K_{0,1},K_{N-1,N+1},K_{N,N+2}$
are respectively the self-energies which renormalise the effect of semi-infinite leads  and the tunneling matrices between the ladder network and leads. 
Noting that all tunneling matrices are unity by scaling in the present calculation, we reach the transmission $T_{j}$ with $j=1,2$ and reflection probabilities $R$,
\begin{eqnarray}
T_{j}&=&\bigl|<N+j|G|0>K^{\ast}_s[F^{-1}(+) \nonumber\\
& &-F^{-1}(-)]\bigr|^2  \qquad (j=1,2),  \nonumber\\
R&=&\bigl|<0|G|0>K^{\ast}_s[F^{-1}(+) \nonumber\\
& &-F^{-1}(-)]-1\bigr|^2
\label{landtr}
\end{eqnarray}

In Fig.~\ref{comp} we compare the results of Eq.~\re{nltr} with those of Eq.~\re{landtr} in case of the ladder with $N=20$. Surprisingly two approaches give the identical results. 
The reason is that the width of the WP employed here is much longer than the linear dimension of the network and that the nonlinearity plays little role. Precisely speaking, so far as  the soliton is large enough and fast enough to guarantee that the  time of collision between the soliton and ladders is much shorter than the soliton dispersion time, one may resort to a linear approximation to compute the transmission coefficients \cite{mir}.
In the following, therefore, we shall derive $T_1,T_2$ and $R$ with use of Eq.~\re{landtr} applied to the linear Schr\"odinger equation for the latter.

\section{Transmission Spectra of Straight ladder}
One cannot recognize any universal feature in Fig. \ref{comp} in the case of a ladder with $n=10$ steps. However, when $n\gg10$, there appear universal characteristic features independent of $n$. 
\begin{figure}[htb]
\begin{center}
\epsfysize=4cm
\epsfxsize=6cm
\epsffile{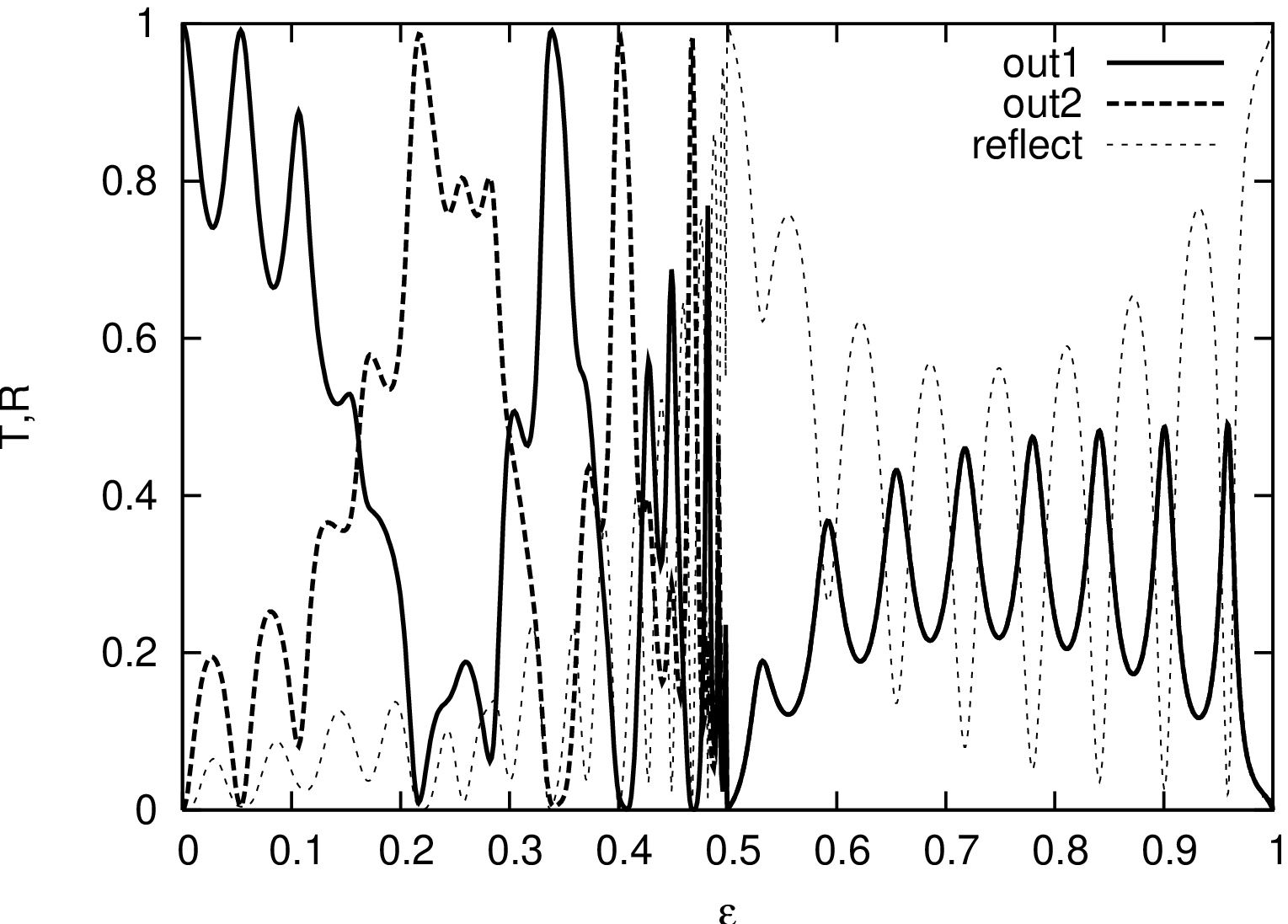}
\epsfysize=4cm
\epsfxsize=6cm
\epsffile{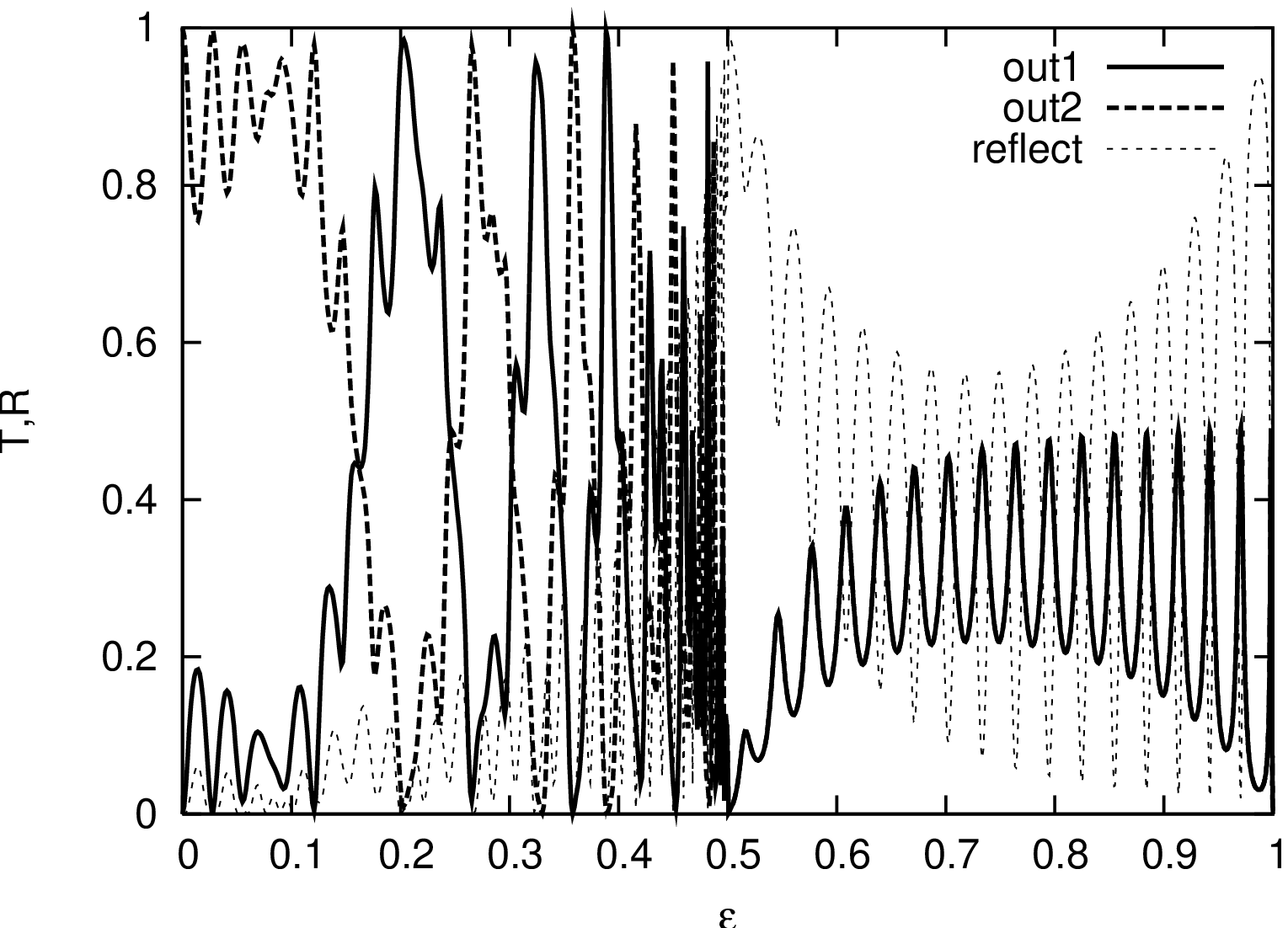}
\epsfysize=4cm
\epsfxsize=6cm
\epsffile{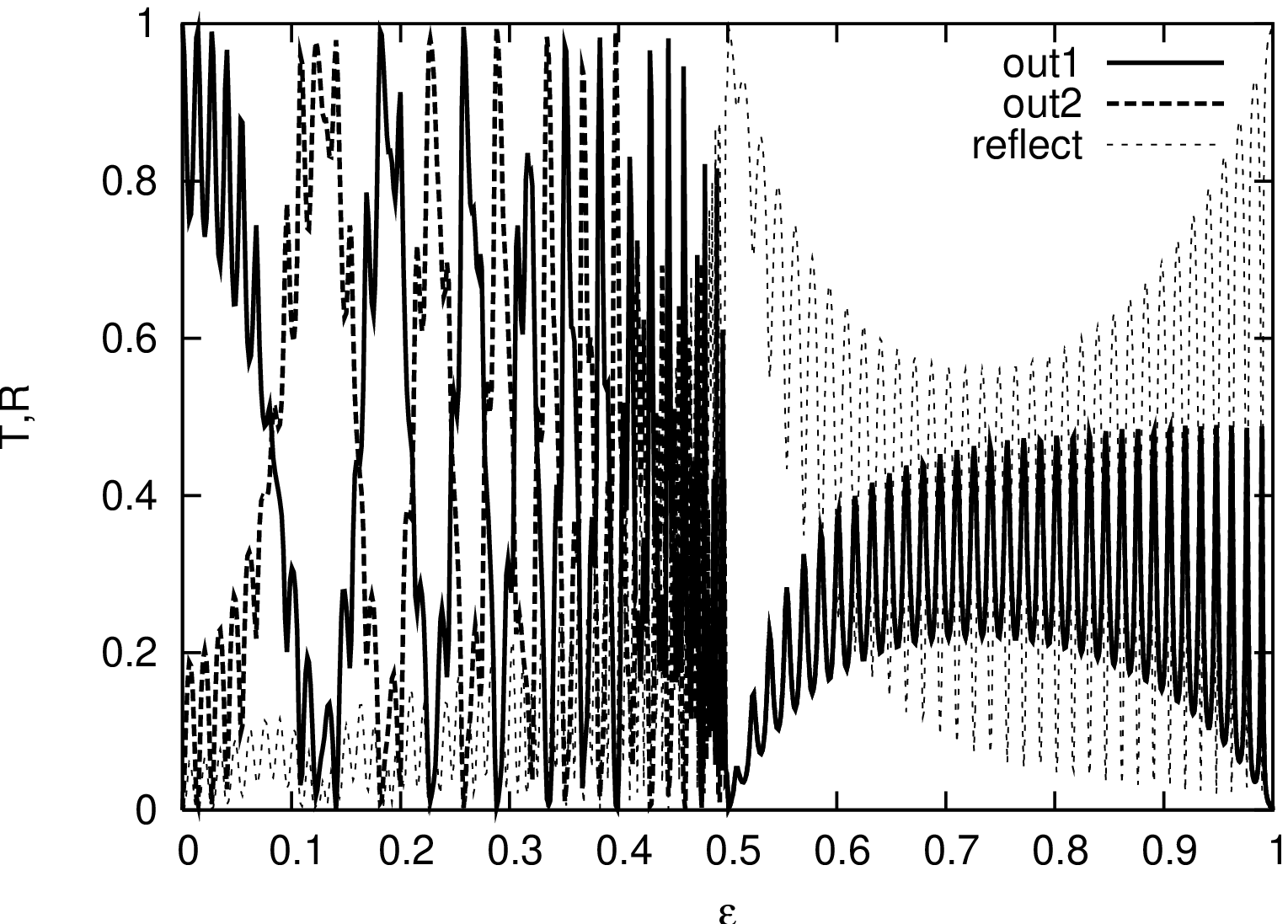}
\caption{Transmission and reflection probabilities  against incident energy $\varepsilon$. Number of steps in the ladder is $n$ = 50, 100 and 200 from top to bottom panels. Solid, dashed and dotted lines correspond to $T_1, T_2$  and $R$, respectively. $T_1$ and $T_2$ are degenerate for $\varepsilon\geq 0.5$. } 
\label{t1t2r}
\end{center}
\end{figure}
In Fig.~\ref{t1t2r} transmission and reflection probabilities against energy ($ \varepsilon$) of the incoming electron are plotted in case of the straight ladder with  $n=50, 100$ and $200$ steps.
The unitarity $T_1+T_2+R=1$ is always satisfied. We find the existence of a critical energy $ \varepsilon_c=0.5 $ and the remarkable difference of TPs between the lower $(0<\varepsilon<\varepsilon_c)$ and higher $(\varepsilon_c<\varepsilon<1)$ energy regions. 
In the lower energy side, $T_1$ and $T_2$ have the anti-phase structure (i.e., $T_1$ takes peaks whenever $T_2$ has dips and vice versa), and the
oscillation period decreases as $\varepsilon\to \varepsilon_c$. In the high energy side, on the other hand, two TPs are degenerate and highly periodic. All these characteristics hold irrespective of the value of $n$, so long as the network is big enough ($n\gg 10$). In fact, we obtained the same 
spectrum in case of $n=1000$ as in Fig.~\ref{t1t2r}, while the oscillation period is further shortened in the latter.

The mechanism underlying the above characteristics is explained by using the perturbation theory. Let's first investigate the nature of the unperturbed long network without three leads, which can be regarded as a periodic ladder  in Fig.~\ref{perf}. 
\begin{figure}[htb]
\centerline{\includegraphics[width=\columnwidth]{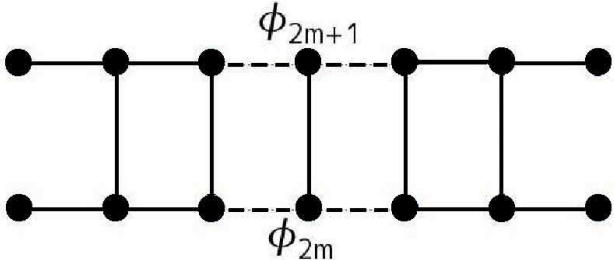}}
\caption{Unperturbed periodic straight ladder.} 
\label{perf}
\end{figure}
For a pair of upper and lower sites $2m$ and $2m+1$,
the wave functions satisfy
\begin{eqnarray}
\varepsilon\Phi_{2m} & = &-\frac{1}{2}(\Phi_{2(m+1)}+\Phi_{2(m-1)}+\Phi_{2m+1}),\nonumber\\
\varepsilon\Phi_{2m+1} & = &-\frac{1}{2}(\Phi_{2(m+1)+1}+\Phi_{2(m-1)+1}+\Phi_{2m}).\label{original upper}
\end{eqnarray}
Let us introduce new basis functions $u_m$ and $v_m$ with use of the transformation:
\begin{eqnarray}
\label{eqper}
\begin{cases} u_m=\frac{1}{\sqrt{2}}(\Phi_{2m}+\Phi_{2m+1}) \\
v_m=\frac{1}{\sqrt{2}}(\Phi_{2m}-\Phi_{2m+1}). \end{cases}
\end{eqnarray}
$u_m$ and $v_m$ stand for the even- and odd-parity states in each step, respectively. Using this new basis, the eigenvalue problem is decoupled, namely, reduced to the even- and odd-parity parts. Assuming $u_m\sim e^{ikm}$ and $v_m\sim e^{ikm}$ for an infinitely long ladder, we find eigenvalues 
\begin{eqnarray}
\varepsilon_u & = & -\cos(k)-\frac{1}{2} \nonumber\\
\varepsilon_v & = &-\cos(k)+\frac{1}{2}.
\label{en}
\end{eqnarray}
\begin{figure}[htb]
\begin{center}
\epsfysize=5.0cm
\epsfxsize=5.0cm
\epsffile{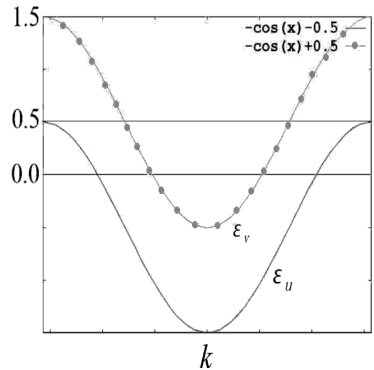}
\caption{Two branches of energy dispersion for unperturbed ladder. Vertical axis stands for energy $\varepsilon$. $\varepsilon_u$: even-parity branch; $\varepsilon_v$: odd-parity branch.} 
\label{disp}
\end{center}
\end{figure}
The even-parity branch $\varepsilon_u$ and odd-parity one $\varepsilon_v$ constitute a pair of energy bands (see Fig.~\ref{disp}).
It should be noted: while for $0\leq\varepsilon\leq\varepsilon_c$, both energy branches $\varepsilon_u$ and $\varepsilon_v$ appear, only the
$\varepsilon_v$ branch can survive for $\varepsilon\geq \varepsilon_c$. 

Under the presence of the perturbation, namely, in the case of the ladder attached with three leads in Fig.~\ref{fig1}, 
$u_m, v_m\sim e^{ikm}$ are not the eigenstates any more: the mixing (superposition) of states occur within
the odd-parity manifold only for $\varepsilon\geq\varepsilon_c$ and between the odd- and even-parity manifolds for
$0\leq\varepsilon\leq\varepsilon_c$. In case of $\varepsilon\ge\varepsilon_c$, therefore, the wave function retains the same feature as the unperturbed state: 
the coefficients of the wave function $\Phi_{2m}$ and $\Phi_{2m+1}$ have the identical magnitude.
This fact holds at the ladder edge with $m=2n$ and $m=2n+1$ as well.
Consequently, we see the degeneracy of oscillations for $T_1$ and $T_2$ in Fig.~\ref{t1t2r}. On the other hand, in case of
$0\leq\varepsilon\leq \varepsilon_c$, we see the superposition of $u_m$ and $v_m$:
\begin{equation}
 \alpha u_m+\beta v_m=\frac{1}{\sqrt{2}}(\alpha+\beta)\Phi_{2m}+\frac{1}{\sqrt{2}}(\alpha-\beta)\Phi_{2m+1}.
\end{equation}
As a result, wherever the coefficient of $\Phi_{2m}$ has a big magnitude, that of $\Phi_{2m+1}$
has a small one, and vice versa. This is true even at the ladder edge, explaining the anti-phase oscillation for $T_1$ and $T_2$ in Fig.~\ref{t1t2r}. 

Thus, the transmission spectra of the straight ladder attached with three leads show a mixing
between different parity states and anti-phase structure in the output in the lower energy regime $(0\le\varepsilon\le\varepsilon_c)$, while, in the higher energy regime $(\varepsilon_c\le\varepsilon\le1)$, no mixing
and the degenerate periodic structure in the output.

\section{Role of defect bonds and topology}
One of the most essential question of quantum networks is whether or not only a single defect bond introduced into big networks will plays a crucial role in quantum transport.
Now we proceed to investigate the influence of  a missing bond embedded in the midst of the  ladder network with $N=100$ steps on the quantum transport.
The left and right panels in Fig.~\ref{mbs} correspond to breaking a bond and step, which are  parallel and perpendicular 
to the ladder, respectively. The corresponding transmission spectra are given in Figs.~\ref{mb} and \ref{ms}.
\begin{figure}[htb]
\begin{center}
\epsfysize=6.0cm
\epsfxsize=6.0cm
\epsffile{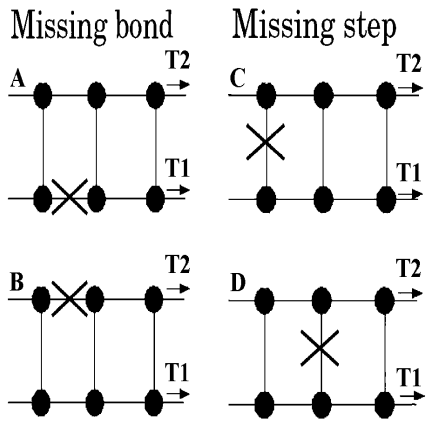}
\caption{Missing bonds $(A,B)$ and missing steps $(C,D)$. Each figure shows only 6 quantum dots in the midst of the long regular ladder. $A (B)$ corresponds the case that a single bond with $\times$, which is parallel to the ladder, is missing.  Missing bond in $(B)$ is displaced upwards from one in $(A)$ by lattice constant;  $C (D)$ corresponds the case that a single step with $\times$, which is perpendicular  to the ladder, is missing. Missing step 
in $(D)$ is displaced to right from one in $(C)$ by lattice constant.} 
\label{mbs}
\end{center}
\end{figure}
\begin{figure}[htb]
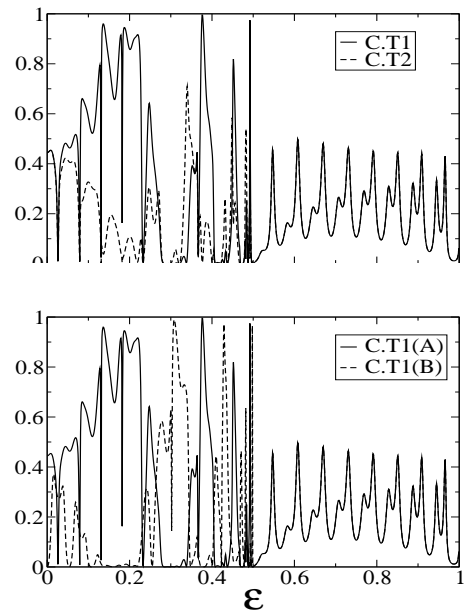

\begin{center}
\epsfysize=4cm
\epsfxsize=6cm
\epsffile{fig10_1.eps}
\epsfysize=4cm
\epsfxsize=6cm
\epsffile{fig10_2.eps}
\caption{Transmission probabilities against incident energy $\varepsilon$ in case of a single missing bond (MB). Numbers of steps ($n$) and of lattice points in the ladder are 100 and 200, respectively. Upper panel includes $T_1$(solid line) and $T_2$(dashed line) in the case that MB lies between lattice points 100 and 102 (: case 'A' in Fig.~\ref{mbs}). Lower panel includes only $T_1$, and solid and dashed lines correspond to cases 'A' and 'B' in Fig.~\ref{mbs}, respectively. Spectra are degenerate for $\varepsilon \geq 0.5$.} 
\label{mb}
\end{center}
\end{figure}
\begin{figure}[htb]
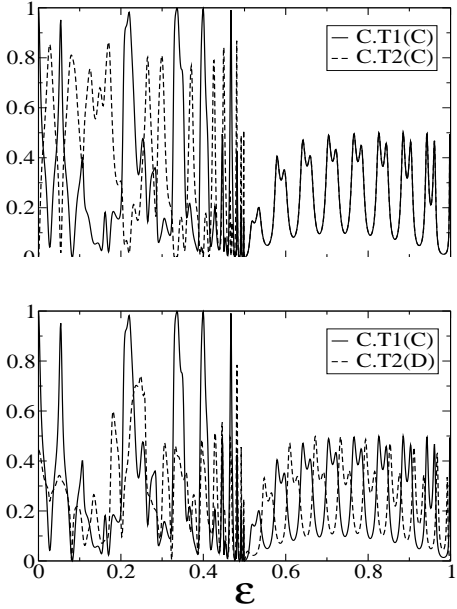

\begin{center}
\epsfysize=4cm
\epsfxsize=6cm
\epsffile{fig11_1.eps}
\epsfysize=4cm
\epsfxsize=6cm
\epsffile{fig11_2.eps}
\caption{The same as Fig. \ref{mb} but in the case of a single missing step (MS). Upper panel includes $T_1$(solid line) and $T_2$(dashed line) in the case that MS lies between lattice points 100 and 101 (: case 'C' in Fig.~\ref{mbs}). Lower panel includes only $T_1$, and solid and dashed lines correspond to cases 'C' and 'D' in Fig.~\ref{mbs}, respectively.} 
\label{ms}
\end{center}
\end{figure}
Consider the case with a missing bond (MB) in the mid-ladder. For $\varepsilon>\varepsilon_c$, the regular oscillation of $T_1$ and $T_2$ retains the degeneracy and in-phase
structure, but has a period twice as large as the one without MB. For $\varepsilon<\varepsilon_c$, $T_1$ shows a radical
change from the complete transmission ($T_1=1$) to the complete reflection ($T_1=0$) and vice versa  when MB moves by lattice constant, which can be taken as a switching effect (see Fig.~\ref{mb}).
\begin{figure}[htb]
\centerline{\includegraphics[width=\columnwidth]{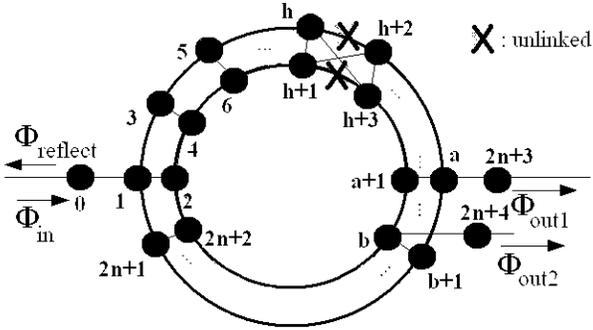}}
\caption{Twisted annular circle. '$\times$' means the disconnection, $h+1$ and $h+2$ (likewise, $h$ and $h+3$) are connected.}
\label{tc}
\end{figure}
\begin{figure}[htb]
\centerline{\includegraphics[width=\columnwidth]{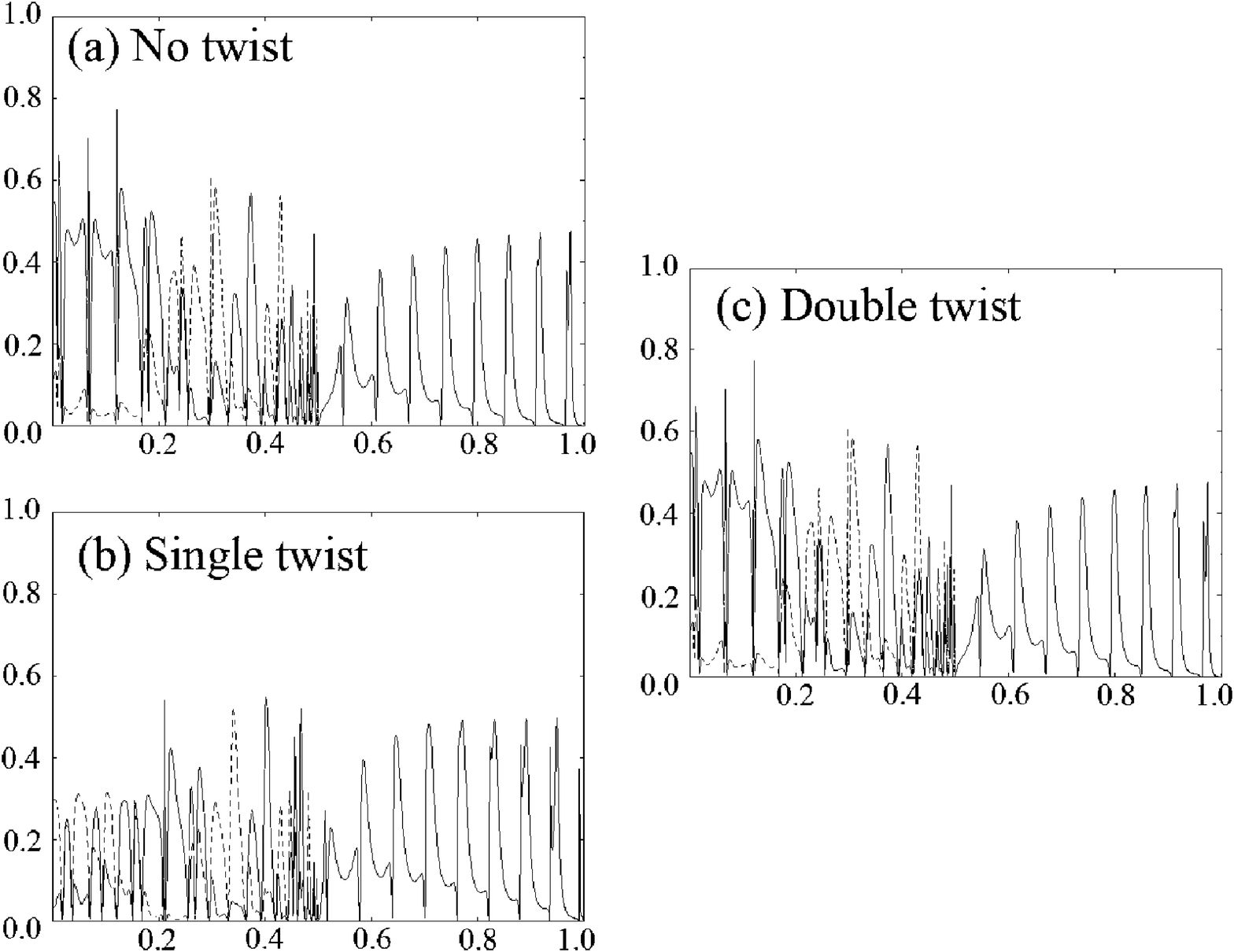}}
\caption{Transmission probabilities against incident energy $\varepsilon$ (solid for $T_1$ and dashed for $T_2$). Three cases of twisted circles: a) no twist; b) a single twist; c) double twists. Spectra are degenerate for $\varepsilon>\varepsilon_c$, though no bold line appears.}
\label{3sp}
\end{figure}
The issue of a missing step (MS) in the midst of the ladder is as follows: for $\varepsilon>\varepsilon_c$, besides the period-doubling phenomenon, the regular oscillation shows a 
phase shift by half a period when MS moves by  lattice constant (see Fig.~\ref{ms}). We should note: so long as a reference MB or MS is embedd in the midst of big networks, the above discoveries (i.e., period doubling and phase shift for $\varepsilon > \varepsilon_c$, and switching effect for $\varepsilon < \varepsilon_c$) remains unchanged, irrespective of the absolute location of such a defect bond in Fig.~\ref{mbs}. Thus, an introduction of a single MB or MS into a big network results in a radical change in the transmission spectra.

In order to see the role of another topology of networks we consider  the annular circular ladder  and investigate the twist effect (see Fig.~\ref{tc}) on quantum transport.

In the case of no twist, the spectra show the same remarkable transition when $\varepsilon$ crosses $\varepsilon_c=0.5$ as in the case of the straight ladder. We find:
In the lower energy side, $T_1$ and $T_2$ have the anti-phase structure, and the
oscillation period decreases as $\varepsilon\to \varepsilon_c$. In the high energy side, on the other hand, two TPs are degenerate and highly periodic.
In the presence of a single twist (i.e., analogue of M\"obius strip) the spectra again shows a remarkable transition at $\varepsilon_c=0.5$, but the detailed feature differs from the result for the no twist case. See the great reduction of $T_1$ and $T_2$ in the lower energy region in the single twist case.  On the other hand, in the double twists case the result is identical to that of no twist case.
The spectra is determined by the parity of the winding number (WN). The winding of the circular ladder is identical to the application of Aharonov-Bohm flux with WN multiplied by a half of the flux quantum $\frac{\phi_0}{2}=\frac{hc}{2e}$. Thus the topology of networks plays a vital role in quantum transport.

\section{Spin-orbit interaction and spin transport}
Recent progress in semiconductor spintronics revealed a way of controlling the magnetization of devices not by a
magnetic but by an electric field. The idea is to use Rashba spin-orbit interaction (SOI) \cite{ras, byc, sou,nik} whose strength is
tuned by the external gate voltage. In this Section, by introducing SOI into the network, we investigate spin
transport (spin-dependent transport) as well as charge transport. According to the pioneering work of Datta and Das 
\cite{datta,dat,zut}, we first consider the spin transport against the spin-polarized injection. The network Hamiltonian generalized so as to include Rashba SOI is given by
\begin{equation}
-\frac{1}{2}\sum_l A_{j,l}\Phi_l+\alpha (\sigma \times p) _z\Phi_j=\varepsilon\Phi_j
\label{Rash}
\end{equation}
with $\Phi_j\equiv(\phi_{j,\uparrow},\phi_{j,\downarrow})^T$ the two component wave function, $\alpha=-\frac{e\hbar}{4m^2c^2K}E_z$ the strength of Rashba SOI in the case of an vertically applied electric field and  $\sigma$ stands for Pauli matrices. In Eq.(\ref{Rash}), energy is scaled by
the tunneling matrix element $K$.
\begin{figure}[htb]
\centerline{\includegraphics[width=\columnwidth]{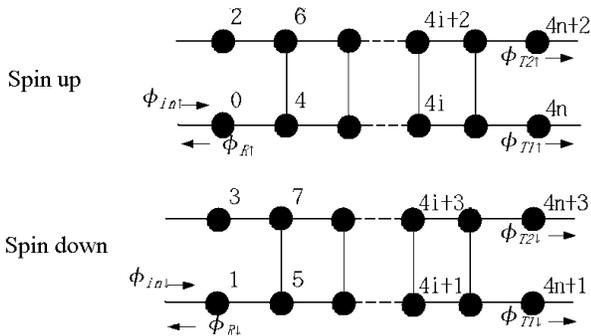}}
\caption{Spinor ladders. For computational purpose, dual ladders are introduced with each corresponding to spin-up and spin-down states.}
\label{spinor}
\end{figure}
\begin{figure}[htb]
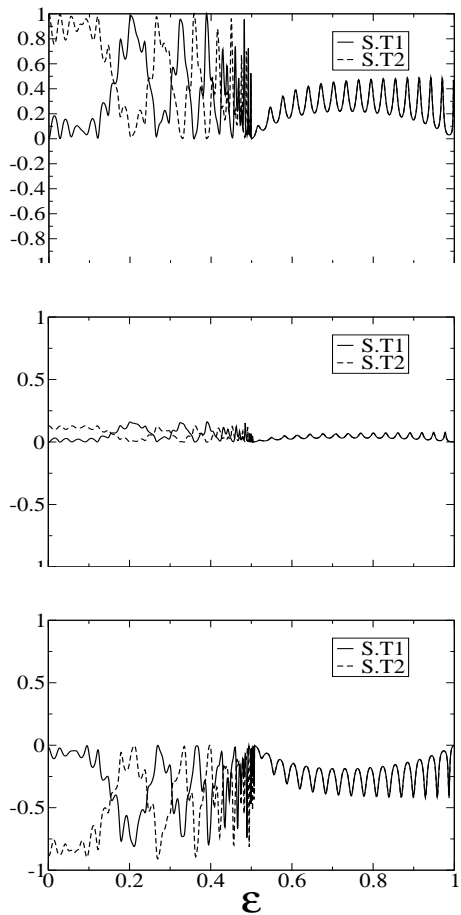

\begin{center}
\epsfysize=4cm
\epsfxsize=6cm
\epsffile{fig15_1.eps}
\epsfysize=4cm
\epsfxsize=6cm
\epsffile{fig15_2.eps}
\epsfysize=4cm
\epsfxsize=6cm
\epsffile{fig15_3.eps}
%\epsfysize=4cm
%\epsfxsize=6cm
%\epsffile{fig15_4.eps}
\caption{Spin transport $T_1^{spin}$ (solid) and $T_2^{spin}$ (dashed) for different values of spin-orbital interaction in case of spin-polarized injection. The panels from top to bottom correspond to 
$\alpha=0.0, 0.12$ and $0.18$, respectively.} 
\label{sptr}
\end{center}
\end{figure}
For convenience in  our numerical calculation, we introduced dual ladders to assign each of them to up- and down-spin states, respectively (see Fig.~\ref{spinor}).
The spin transport is quantified as $T_{1,2}^{spin}=T_{1,2}(\uparrow)-T_{1,2}(\downarrow)$ and the charge transport as
$T_{1,2}^{charge}=T_{1,2}(\uparrow)+T_{1,2}(\downarrow)$.

In Fig.~\ref{sptr} the spin transport against incident energy is plotted for different values of the strength of Rashba spin-orbit interaction $\alpha$. We consider the spin-polarized $(S_z=+\frac{1}{2})$ injection. In the absence of spin-orbital interaction the spin transport (STP)  $T_1^{spin}, T_2^{spin}$ as a function of $\varepsilon$  show the same spectra  as in the case of charge transport $T_1, T_2$
(see Fig.~\ref{t1t2r}), because we have no contribution from $T_{1,2}(\downarrow)$.
Against the variation of SOI, the spin transport shows spin-flip (magnetization reversal) oscillations (see Fig.~\ref{sptr}), while keeping the anti-phase structure of $T_{1}^{spin}$ and $T_{2}^{spin}$ in the range $\varepsilon<\varepsilon_c(=0.5)$. Against the variation of SOI, by contrast, the charge transport (CTP) keeps the spectral feature without SOI (see Fig.~\ref{t1t2r}).

Finally we shall investigate the most interesting subject, namely the spin transport in network systems with SOI against the injection of spin-unpolarized electron.  Figure \ref{sptr2} shows $T_1^{spin}$ and $T_2^{spin}$ as a function of 
$\varepsilon$ for non-zero values of $\alpha$. Astonishingly we find $T_1^{spin} =-T_2^{spin}$ for any value of $\varepsilon$
in the case of $\alpha \neq 0$. This discovery indicates that a straight ladder with three leads plays a role of the spin filtering, i.e., the unpolarized electron is decomposed into mostly spin-up and mostly spin-down components through its transport in the ladder. In the context of nanoscience,
this is the most essential issue among many other discoveries in the present work.
\begin{figure}[htb]
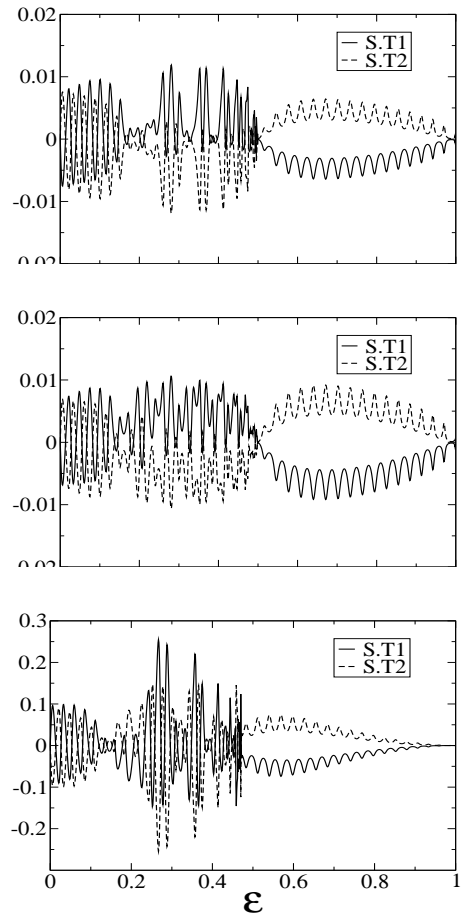

\begin{center}
\epsfysize=4cm
\epsfxsize=6cm
\epsffile{fig16_1.eps}
\epsfysize=4cm
\epsfxsize=6cm
\epsffile{fig16_2.eps}
\epsfysize=4cm
\epsfxsize=6cm
\epsffile{fig16_3.eps}
\caption{The same spin transport $T_1^{spin}$ (solid) and $T_2^{spin}$ (dashed) as in Fig. 
\ref{sptr}, but in the case of spin-unpolarized injection. The panels from top to bottom correspond to $\alpha=0.1, 0.12$ and $0.46$, respectively.} 
\label{sptr2}
\end{center}
\end{figure}
\section{Summary and discussions}

Choosing straight and circular ladders as big network models and attaching them with one incoming and  two outgoing semi-infinite leads, 
we examined quantum transport of an electron or phase soliton.
In the beginning, by adding a small cubic nonlinearity (e.g., Hartree term) to the discrete time-dependent linear Schr\"odinger equation, we showed how the
incoming soliton bifurcates at the entrance of the ladder-type network and is ultimately evacuated from the network through  three
leads. 

We chose a soliton large enough and fast enough to guarantee the time of collision between the soliton and ladders to be much shorter than the soliton dispersion time. On the basis of this soliton picture, two transmission probabilities ($T_{1,2}$) and a reflection probability ($R$) were evaluated, which proved to accord with the corresponding probabilities obtained from the linear methodology, i.e., Landauer formula applied to the time-independent linear Schr\"odinger equation. The main part of the paper was then  devoted to the results of the latter (linear) methodology.
Firstly we investigated $T_1$, $T_2$ as a function of energy $\varepsilon$ of the incident electron. Both probabilities show a
transition from anti-phase aperiodic to degenerate periodic spectra at the critical energy $\varepsilon_c=0.5$, whose value  is determined
by a bifurcation point of the bulk energy dispersions.  TPs  of the circular ladder depend only on the parity of the winding number (WN), because WN plays a role of Aharonov-Bohm flux with its magnitude being a half of flux quantum multiplied by WN.

Introduction of a single defect bond into big networks radically changes the macroscopic transport spectra. A missing bond (MB) parallel to the ladder in the network doubles  period of the periodic spectra for 
$\varepsilon>\varepsilon_c$.  For $\varepsilon<\varepsilon_c$, shift of a single MB by lattice constant results 
in the switching between two outgoing leads. A missing step leads to a phase shift besides the period doubling for $\varepsilon>\varepsilon_c$.

Finally, by introducing the electric-field-induced Rashba spin-orbit interaction (SOI), we explored  spin transport ($T_1^{spin}$, $T_2^{spin}$) against the spin-polarized injection. 
At zero SOI, $T_1^{spin}$ and $T_2^{spin}$ as a function of $\varepsilon$  show the same spectra  as in the case of charge 
transport. Against a variation of SOI, however,  this structure shows a coherent spin-flip (magnetization reversal) oscillations. On the other hand, the injection of the spin-unpolarized electron leads to the spin filtering, namely, the unpolarized electron is decomposed spatially into mostly spin-up and mostly spin-down components through its transport in the ladder. Therefore the present network can be used as a spin-filtering device. This is the most striking issue of this paper.
The present results would also be applicable
to propagation of a wide-enough soliton in Josephson junction networks and of a wave packet in Bose-Einstein condensates in optical-lattice networks, although the linear and static approximation will break down and the transport would be highly nonlinear and more generic.

{\bf Acknowledgment}

We are grateful for valuable discussions with A. Terai, Y. S. Kivshar, B. Abdullaev, and  F. Abdullaev. The work is partly supported through a project of the Uzbek Academy of Sciences (FA-F2-084).

%\newpage

\end{document}